\documentclass[12pt]{article}
\usepackage{graphicx}
\usepackage{amsmath}
\usepackage{amssymb}
\usepackage{subfigure}
\usepackage{latexsym}
\usepackage{hyperref}
\usepackage[sort]{cite}

\newcommand{\be}{\begin{eqnarray}}
\newcommand{\ee}{\end{eqnarray}}
\newcommand{\bc}{\begin{center}}
\newcommand{\ec}{\end{center}}
\newcommand{\bea}{\begin{eqnarray}}
\newcommand{\eea}{\end{eqnarray}}
\newcommand{\ben}{\begin{equation}}

\newcommand{\del}{\partial}

\newcommand{\nn}{\nonumber}

\numberwithin{equation}{section}

\newsavebox{\ns}
\newsavebox{\dbrane}
\newsavebox{\dbshort}

\usepackage{epsfig}
\usepackage{amssymb}

\def\appendix{{\newpage\section*{Appendix}}\let\appendix\section%
        {\setcounter{section}{0}
        \gdef\thesection{\Alph{section}}}\section}

\newcommand\ba{\begin{eqnarray}}
\newcommand\ea{\end{eqnarray}}

\def\Dslash{\,\,{\raise.15ex\hbox{/}\mkern-12mu D}}
\def\Dbarslash{\,\,{\raise.15ex\hbox{/}\mkern-12mu {\bar D}}}
\def\delslash{\,\,{\raise.15ex\hbox{/}\mkern-9mu \partial}}
\def\delbarslash{\,\,{\raise.15ex\hbox{/}\mkern-9mu {\bar\partial}}}
\def\pslash{\,\,{\raise.15ex\hbox{/}\mkern-9mu p}}
\def\calDslash{\,\,{\raise.15ex\hbox{/}\mkern-12mu {\cal D}}}

\newcommand{\hh}{{1\over 2}}

\renewcommand{\ll}{_}
\newcommand{\uu}{^}
\newcommand{\pp}{\partial}

\renewcommand{\exp}[1]{{\rm exp}\left(#1\right)}

\newcommand{\m}{\mu}

\renewcommand{\m}{\mu}
\newcommand{\n}{\nu}
\newcommand{\s}{\sigma}
\renewcommand{\t}{\tau}
\newcommand{\G}{\Gamma}
\newcommand{\g}{\gamma}
\renewcommand{\a}{\alpha}
\renewcommand{\r}{\rho}

\renewcommand{\o}{\omega}

\renewcommand{\O}{\Omega}

\newcommand{\sqd}{^2}

\renewcommand{\hh}{{1\over 2}}

\newcommand{\eee}[1]{\ba{#1}\ea}

\renewcommand{\t}{\tau}

\renewcommand{\b}{\beta}

\newcommand{\pr}{^\prime {}}

\newcommand{\apr}{{\alpha^\prime} {}}

\newcommand{\IZ}{\relax\ifmmode\mathchoice
{\hbox{\cmss Z\kern-.4em Z}}{\hbox{\cmss Z\kern-.4em Z}}
{\lower.9pt\hbox{\cmsss Z\kern-.4em Z}} {\lower1.2pt\hbox{\cmsss
Z\kern-.4em Z}}\else{\cmss Z\kern-.4em Z}\fi} \font\cmss=cmss10
\font\cmsss=cmss10 at 7pt
\newcommand{\inbar}{\,\vrule height1.5ex width.4pt depth0pt}
\newcommand{\IC}{{\relax\hbox{$\inbar\kern-.3em{\rm C}$}}}
\newcommand{\IQ}{{\relax\hbox{$\inbar\kern-.3em{\rm Q}$}}}
\newcommand{\IP}{\relax{\rm I\kern-.18em P}}

\renewcommand{\k}[1]{{k_{#1}}}

\newcommand{\ed}{\dot{e}}

\renewcommand{\ge}{{\tilde{\e}}}

\renewcommand{\k}{\kappa}

\newcommand{\phd}{\dot{\phi}}

\renewcommand{\o}{\omega}

\newcommand{\ct}{\tilde{c}}

\renewcommand{\pr}{{}^\prime{}}

\newcommand{\IR}{\relax{\rm I\kern-.18em R}}
\def\blfootnote{\xdef\@thefnmark{}\@footnotetext}

\newcommand{\bm}{\begin{matrix}}

\newcommand{\lba}{\left |}
\newcommand{\rba}{\right |}

\newcommand{\rr}[1]{(\ref{{#1}})}
\newcommand{\bbb}{\ba}
\renewcommand{\eee}{\ea}
\newcommand{\een}[1]{\label{#1}\ea}

\newcommand{\Xd}{{\dot{X}}}

\def\hilo{{}_{{}_{{}_{{}_{{}_{}}}}} {}^{{}^{{}^{}}}}

\newcommand{\heading}[1]{\begin{center}\it {#1} \rm \end{center}}

\def\lrdd{\left ( ~}
\def\rrdd{\hilo \right )}
\def\lsqq{\left [ ~}
\def\rsqq{\hilo \right ]}

\newcommand{\us}[2]{^{({#1}{#2})}}

\def\bi{\begin{itemize}}
\def\ei{\end{itemize}}

\def\ed{\end{document}}

\def\ge{{G^{(E)}}}
\def\gs{{G^{(S)}}}

\def\ue{^{(E)}}

\def\us{^{(S)}}
\renewcommand{\rr}[1]{(\ref{#1})}
\def\tb{\bar{t}}
\def\marker{\lambda}
\def\momp{p}
\def\newv{\cal V}
\def\vertexoperator{\cal U}
\begin{document}

\begin{titlepage}
\begin{flushright}
hep-th/0611317
\end{flushright}
\vspace{15 mm}
\begin{center}
{\Large \bf Cosmological solutions of supercritical string theory }
\end{center}
\vspace{6 mm}
\begin{center}
{ Simeon Hellerman and Ian Swanson }\\
\vspace{6mm}
{\it School of Natural Sciences, Institute for Advanced Study\\
Princeton, NJ 08540, USA }
\end{center}
\vspace{6 mm}
\begin{center}
{\large Abstract}
\end{center}
\noindent
We study quintessence-driven, spatially flat, expanding FRW
cosmologies that arise naturally
from string theory formulated in a supercritical number of spacetime dimensions. 
The tree-level potential of the string
theory produces an equation of state at the threshold 
between accelerating and decelerating cosmologies, and the resulting
spacetime is globally conformally equivalent to Minkowski space.
We demonstrate that exact solutions exist with a
condensate of the closed-string tachyon, the simplest of 
which is a Liouville wall moving at the speed of light. 
We rely on the existence of this solution
to derive constraints on the couplings of the tachyon to the dilaton 
and metric in the string theory effective action.  
In particular, we show that the tachyon dependence of the Einstein term must
be nontrivial. 
\vspace{3cm}
\vspace{1cm}
\begin{flushleft}
November 29, 2006
\end{flushleft}
\end{titlepage}

\section{Introduction}
String theory formulated in a supercritical number of spacetime dimensions 
provides a natural setting for the study of time-dependent backgrounds
in gravitational physics.    The simplest classical solutions of supercritical string 
theory exhibit a string-frame metric
that is flat and a dilaton that has a timelike gradient
proportional to $\sqrt{D - D\ll{\rm crit}}$, where $D_{\rm crit}$ is the critical number of 
spacetime dimensions of the theory: 26 for the bosonic string or 10 for the superstring.
In this paper we describe these theories in the language of an expanding 
FRW cosmology driven by quintessence.
(Other studies along these lines can be found in
\cite{Tseytlin:1991bu,Tseytlin:1991xk,Tseytlin:1992ye,Tseytlin:1992jq,Diamandis:2001nn,Diamandis:2002zn,Antoniadis:1990uu,Antoniadis:1988vi,Antoniadis:1988aa}, for example.  
Recent work also includes \cite{Ellis1,Ellis2}.)

Quintessent cosmologies are defined by an equation of 
state $w \equiv p/\r$ taking some fractional negative value not less than $-1$.  
(Situations with $w < -1$
are forbidden by the null energy condition.)  
In four dimensions, an interesting range of state equations 
for cosmologists to consider is $-1 \leq w < - 1/3$:  in this range
the scale factor $a(t)$
of the FRW universe accelerates as a function of the proper time
$t$ measured 
by static observers (we refer to this as FRW time).  The simplest way to generate models of 
quintessence is to adopt a matter system described by a single, real scalar
field entering the action with a canonical kinetic term and an exponential potential,
\be
{\newv}(\phi ) = c\, \exp{\g \phi }\ ,
\ee
with positive coefficient $c > 0$.  
The value of $\g$ in the exponent determines the equation of state 
$w$ for the theory.

String theory provides several ways of obtaining exponential 
potentials for scalar fields.  The most straightforward
is via the tree-level potential for the dilaton that
arises in string theories formulated in $D\neq D\ll{\rm crit}$ spacetime 
dimensions.  In particular, $c$ is positive for $D>D\ll{\rm crit}$.\footnote{For 
$D = D\ll{\rm crit}$, the tree-level potential vanishes in the
absence of flux or curvature, while for $D< D\ll{\rm crit}$, the sign of this 
potential is negative and the FRW constraint equation cannot be satisfied for 
a spatially flat universe.}  
The resulting theories are characterized as timelike linear dilaton backgrounds,
and they exhibit an extremely simple set of quintessent 
cosmologies that can be analyzed using worldsheet techniques.\footnote{One of the conclusions 
of \cite{hks} was that the
absence of well-defined observables at late times can cause trouble for the interpretation
of cosmologies 
characterized by $w < w\ll{\rm crit}$.
However, backgrounds that asymptote to timelike linear dilaton theories
exhibit well-defined final states described by
a Hilbert space of free particles.  }
We should note, however, that the point here is {\it not } to suggest that
the timelike linear dilaton theory in $D$ dimensions is, by
itself, a phenomenologically accurate model of cosmology.
We merely intend to demonstrate that string theory can generate simple time-dependent 
models of rolling massless scalars with flat potentials.

In our next section we study the timelike linear
dilaton background in $D$ dimensions from a cosmological perspective:
this analysis, in essence, generalizes the presentation in \cite{hks} to
arbitrary dimension.  We demonstrate that 
the critical equation of state $w\ll{\rm crit}$, above which the universe
ceases to accelerate, depends on the number of background dimensions $D$. 
In turn, the global causal structure of the spacetime depends 
on the value of $w$ relative to $w\ll{\rm crit}$. 
We also introduce the spacetime effective action and 
analyze the physics of small fluctuations about
a quintessent background, showing that the cosmology
is stable under perturbations of massless modes.
This section is intended to describe many established facts
about the supercritical string in a framework that we hope may 
be useful to cosmologists.

In Section 3 we turn to a worldsheet analysis of the 
supercritical string.  The quintessent cosmological background is
unstable against perturbations of a tachyon,\footnote{In a generic 
number of dimensions, tachyon-free timelike
linear dilaton theories do not exist.}
analogous to the Poincar\'e-invariant background of the 
critical string.   We demonstrate that, for the supercritical string,
there are certain exact solutions for which the tachyon 
is non-zero.  The simplest of these solutions is a `bubble of nothing' that
destroys the spacetime from within, exhibiting behavior similar to 
that of the Witten solution of \cite{wittenbubble}.  (Other
connections between bulk closed-string tachyon condensation
and the Witten instanton have been proposed in 
\cite{Horowitz:2005vp,Hirano:2005sg,Headrick:2004hz,Gutperle:2002bp,Emparan:2001gm,Costa:2000nw,Adams:2005rb,Aharony:2002cx,zamolunpub}.)
Relying on worldsheet techniques, we show that these solutions
are in fact exact to all orders, and
nonperturbatively, in $\apr$.  The existence of such solutions
implies constraints on the terms appearing in the effective action, 
and consistency demands nontrivial tachyon dependence of the Einstein 
term.
In the final section we discuss our conclusions and describe how our
work might be extended.  

Upon completion of this work, we learned of a paper in preparation
\cite{evaofer} that addresses issues related to supercritical string theory.  Specifically,
the authors of \cite{evaofer} analyze the spacetime physics of the timelike linear dilaton
background as a quintessent FRW cosmology, and come to conclusions consistent with our own.

\section{Timelike linear dilaton theories as quintessent cosmology}
In this section we study the timelike linear dilaton background in standard cosmological 
language. While this section contains no new results,
it should serve as a useful review.

Timelike linear dilaton theories are really
quintessent cosmologies.  This observation is typically obscured by the fact
that the string sigma model is naturally described in terms of the 
string-frame metric rather than the Einstein metric (the two are related
by a conformal transformation).  The Einstein metric $G\ll{\m\n}\ue$ is, 
by definition, the metric whose Ricci scalar
$R^{(E)}$ appears multiplied by a field-independent coefficient in the action:
\bbb
{\cal L} \sim {1\over{2\k\sqd}} \sqrt{-\det{\ge}}~R^{(E)} \ .
\eee
(We will use the labels $(S)$ and $(E)$ to denote string and Einstein frame, 
respectively, and the gravitational coupling $\k$ is related to the Newton
constant by $G\ll N = {{\k\sqd}\over{8\pi}}$.)
Alternatively, the string-frame metric $G\ll{\m\n}\us$ is the metric 
whose Ricci scalar $R\us$ enters the action multiplied by an exponential
of the dilaton $\Phi$:
\bbb
{\cal L} \sim {1\over{2\k\sqd}} 
\exp{- 2  \Phi}
\sqrt{-\det{\gs}}~R\us\ .
\eee
The precise relation between the two constitutes a Weyl transformation 
depending on the dilaton:
\bbb
G\ll{\m\n}\us = \exp{{{4\Phi}\over{D - 2}}} G\ll{\m\n} \ue\ .
\een{weyl}
In this section we aim to study such theories, paying
particular attention to the dependence of various cosmological properties 
on the number of spacetime dimensions.
We will restrict attention to
cases for which homogeneous spacelike hypersurfaces are
flat (in other words, cases for which $k = 0$).  
Among other reasons, this allows us to adopt models of the universe that
we may assume have undergone significant periods of inflation.

\subsection{Quintessent cosmologies with scalar fields}
We first solve the equations of motion in quintessent 
cosmological backgrounds in $D$ spacetime dimensions.
For this, one may closely follow the treatment in \cite{hks}.
The metric for a spatially flat ($k = 0$) FRW cosmology is
\bbb
ds\sqd = - dt\sqd + a(t)\sqd dx\uu i dx\uu i\ , 
\eee
where $i = 1,\cdots, D-1$.  With the definition $w\equiv p / \r$,
the equation of motion for the scale factor $a(t)$ is
\bbb
{{\ddot{a}}\over a} = -
{{D - 3 + w (D - 1)}\over{(D-1)(D-2)}} \kappa^2 \r \ .
\eee
(At this stage we are not assuming that $w$ is independent of time.)
The constraint equation appears as
\bbb
H\sqd = {2\over{(D - 1)(D - 2)}} \k\sqd \r\ ,
\eee
where $H$ is the Hubble constant $H\equiv \dot{a} / a$.  
The matter content of the theory is taken to be a real scalar field
$\phi$ with Lagrangian
\bbb
{\cal L}_\phi = {1\over{\k\sqd}} 
\sqrt{- \det G\ue}
\lsqq - \hh ( \pp\ll \m \phi)(\pp\uu\m\phi) - {\newv}(\phi) \rsqq\ ,
\een{quaction}
where
\bbb
{\newv}(\phi)\equiv c\, \exp{\g \phi},  \qquad c, \g > 0\ .
\eee
This yields the following stress tensor (respectively, the energy density 
and pressure):
\bbb
\rho \equiv {1\over{\k\sqd}} \left(\hh \phd\sqd + {\newv} \right)\ ,
\een{important}
\bbb
p =  {1\over{\k\sqd}} \left(\hh \phd\sqd - {\newv} \right)\ .
\een{important2}
With this input, the constraint equation becomes
\bbb
\frac{\dot\phi^2}{2} + {\newv} = \frac{1}{2}(D-2)(D-1)H^2\ .
\eee
We assume here
that the Einstein term is to remain canonical
\bbb
{\cal L}_{\rm Einstein} = {1\over{2\k\sqd}} 
\sqrt{- \det G\ue}
R\ue\ ,
\een{einac}
and, if the scalar field respects the $(D-1)$-dimensional 
Poincar\'e symmetries of the $k=0$ spatial slices, 
the equation of motion for $\phi$ becomes
\bbb
\ddot{\phi} + (D-1) H \dot \phi = - {\newv}\pr(\phi)\ .
\een{important3}

At this point, we adopt the ansatz that our solution exhibits a constant
equation of state $w$.  It follows that $\phd\sqd,~H\sqd$
and ${\newv}$ all scale as $t\uu{-2}$, so we find the general expressions
\bbb
\phi(t) &=& \marker \log (t / t\ll 1)\ ,
\nn\\ \nn\\
a(t) &=& a\ll 0 \lrdd {t\over{t\ll 0}} \rrdd\uu \a\ ,
\eee
for some $\a, \marker$.
From the fact that ${\newv}$ scales as $c (t\ll 1 / t)\sqd$, we
conclude that
\bbb
\marker \g &=& - 2\ .
\eee
Subjecting ${\newv}$ (as a function of time) to the
restriction of constant equation of state yields
\bbb 
ct\ll 1 \sqd = {2\over{\g\sqd}} {{1 - w}\over{1 + w}}\ .
\eee
It follows that the constraint equation implies
\bbb
\hh (D - 1)(D - 2)\a\sqd = c t\ll 1 \sqd + {2\over\g\sqd}\ .
\eee
To solve for $\a$,
we differentiate Eqn.~\rr{important} with respect to
time and use Eqns.~\rr{important3} and \rr{important2}
to determine
\bbb
{\dot{\rho}} +  {{\dot{a}}\over a} (D-1) (1 + w) \r = 0\ .
\eee
This can be integrated to yield
\bbb
\r = \r\ll 0 \lrdd {{a\ll 0}\over a}\rrdd\uu{{(D - 1)(1 + w)}}\ .
\een{powerlaw}
To fix $\r\ll 0$, we evaluate the 
constraint equation at $t\ll 0$ and require
that $\r =  \r\ll 0$ at time $t\ll 0$, which gives
\be
\r\ll 0 = {4\over{(D-1) t\ll 0\sqd (1 + w)\sqd \k\sqd}}\ .
\ee
Finally, we substitute Eqn.~\rr{powerlaw} into the constraint equation.
This amounts to a direct relation between $\dot{a}$ and $a$, 
which can again be integrated to yield the following:
\be
\a &\equiv& {2\over{(1 + w) (D - 1)}}\ ,
\nn\\ \nn\\
\g\sqd &=& {{2 (D - 1) (w + 1)}\over{D - 2}}\ ,
\nn\\ \nn\\
w &=& -1 + {{(D - 2) \g\sqd}\over{2 (D - 1)}}\ .
\een{wformula}

Because $c>0$, the energy density $\r$ is positive-definite, and the
cosmological scale $a$ accelerates as a function of FRW time if (and
only if) $-1 \leq w < w\ll{\rm crit}$, where
\bbb
w\ll{\rm crit} = - {{D - 3}\over{D - 1}}\ ,
\eee
in $D$ spacetime dimensions.


The global causal structure of the solution depends on whether
$w$ is less than, greater than, or equal to the critical value $w\ll{\rm crit}$.
The spatial slice $t = 0$ defines a singularity in all three cases.
The precise nature of this singularity and the nature of the asymptotic 
future $t\to + \infty$, however, depend on the state equation of the
cosmology.  

To investigate these issues in greater detail, 
we begin by demonstrating the coordinate transformation that
puts the metric in canonical form for a conformally flat spacetime.
In FRW coordinates, we have
\bbb
ds\sqd = - dt\sqd + a\ll 0 \sqd \lrdd {{t\over{t\ll 0}}} \rrdd\uu
{{ 4 / (1 + w) (D - 1)}} dx\uu i dx\uu i\ . 
\eee
Working by analogy from \cite{hks}, we define a new time coordinate $\tb$ via the equation
\bbb
\tb \equiv 
\lrdd
{{(D - 1)(1 + w)}
\over{(D - 1) w + (D - 3)}}
 t\ll 0 \uu{ { 2\over{(D - 1)(1 + w)}} }
 a\ll 0 \uu{-1} \rrdd
t\uu{{(D-1)w + (D -3)}\over{{(D - 1)(1 + w)}}}\ .
\een{tbardef}
In these coordinates, the metric takes the form
\bbb
ds\sqd = \o(\tb)\sqd \lsqq - d\tb \sqd + dx\uu i dx\uu i \rsqq
= \o(\tb)\sqd \lsqq - d\tb\sqd + dr\sqd + r\uu{D - 2} d\O\ll{D - 2}\sqd \rsqq\ ,
\eee
where we have defined
\be
\o(\tb) &\equiv& l\, 
\left( {{(D - 1) w + (D - 3) }\over{(D - 1)(1 + w)}} 
~\tb
\right)\uu{{2\over{(D - 1)w + (D - 3)}}}\ ,
\nn\\ \nn\\
l &\equiv & a\ll 0 \lrdd {{a\ll 0}\over{t\ll 0 }} \rrdd\uu{2\over{(D - 1) w
+ (D - 3)}}\ .
\ee

In an accelerating universe, the combination $${{(D - 1)(1 + w)}
\over{(D - 1) w + (D - 3)}}$$ is negative, and the range of $\tb$
is $\tb \in (-\infty, 0)$.  The initial singularity occurs at $\tb = - \infty$,
and the infinite future lies at $\tb = 0$.  
Alternatively, the combination 
above is positive for a decelerating cosmology, 
and $\tb$ spans the range $\tb \in (0, \infty)$.  
In this latter case the initial singularity occurs at $\tb = 0$,
and the infinite future is located at $\tb = + \infty$. 

For the sake of completeness, we will briefly
summarize the structures of the Penrose diagrams appearing for the various ranges of $w$ described
above.  To construct these diagrams, one ignores 
the $(D-2)$-sphere fibered over each diagram and conformally compactifies 
the $(r,t)$ plane using the transformation 
\be
r \equiv  \frac{\sin \chi}{\cos\chi + \cos\tau}\ , 
\qquad 
\tb \equiv \frac{\sin \tau}{\cos\chi + \cos\tau}\ .
\label{confcomp}
\ee
In these coordinates, the metric on the $(r,t)$ plane becomes
\bbb
ds\sqd = {\frac{l\sqd}{4}} 
{
{
{\left
( \sin |\t| \right)\uu{2\Delta}}
}
\over
{
\left[ \cos\left( {{\chi + \t}\over 2} \right)
 \cos\left( {{\chi - \t}\over 2} \right)\right]
\uu{{ 2 + 2 \Delta}}
}
}
\lba \hilo {{(D-1)w + (D- 3)}\over{2 (D - 1) (1 + w)}} 
\rba  \uu{{2 \Delta }}
{( - d\t\sqd + d\chi\sqd )}\ ,
\een{penrosemetric}
with
\be
\Delta \equiv \frac{2}{(D - 1) w + (D - 3)}\ .
\ee
For $-1 < w < w\ll{\rm crit}$, the constant $\Delta$
is negative and the range of the $\tau$ and $\chi$ coordinates is
\bbb
\t\in [-\pi,0]\ ,
\qquad
\chi \in [0,\t + \pi]\ , \qquad {\rm (accelerating\ universe).}
\eee
For $w > w\ll{\rm crit}$, the quantity $\Delta$
becomes positive and we have
\bbb
\t\in [0,\pi]\ ,
\qquad
\chi \in [0, \pi - \t]\ , \qquad {\rm (decelerating\ universe).}
\eee

\begin{figure}[htb]
\begin{center}
\includegraphics[width=4.0in,height=3.5in,angle=0]{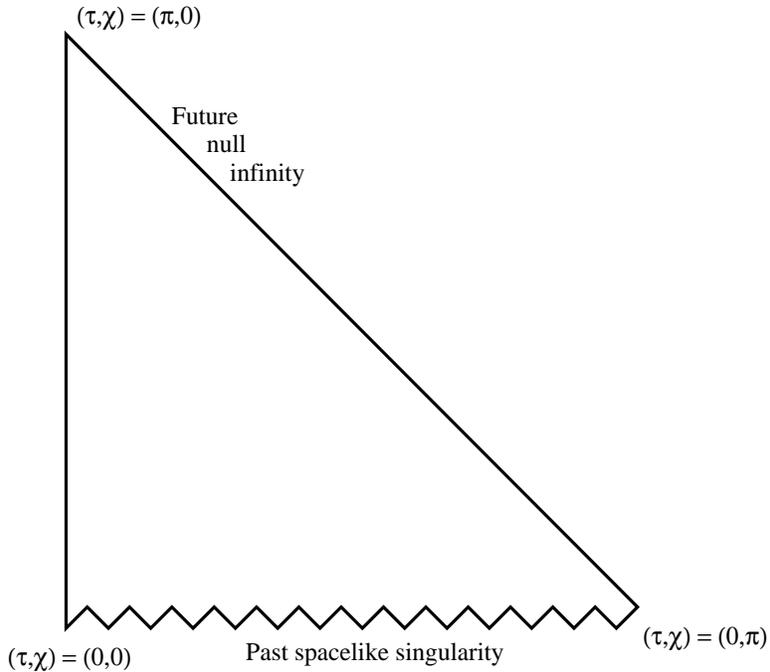}
\caption{Penrose diagram of the decelerating universe ($w > w\ll{\rm crit}$).
The initial singularity is spacelike, and the future boundary is null.}
\label{penrose2}
\end{center}
\end{figure}

In the decelerating case, the state equation lies in the range $w > w\ll{\rm crit}$, and 
it is easy to see that the spatial slice $t=0$ is an ordinary, spacelike Big-Bang singularity 
with divergent scalar curvature (for $w \neq  -1$).  In addition, future infinity 
is null, as it is in ordinary Minkowski space.  This structure is depicted in detail 
in Fig.~\ref{penrose2}.

\begin{figure}[htb]
\begin{center}
\includegraphics[width=3.0in,height=3.0in,angle=0]{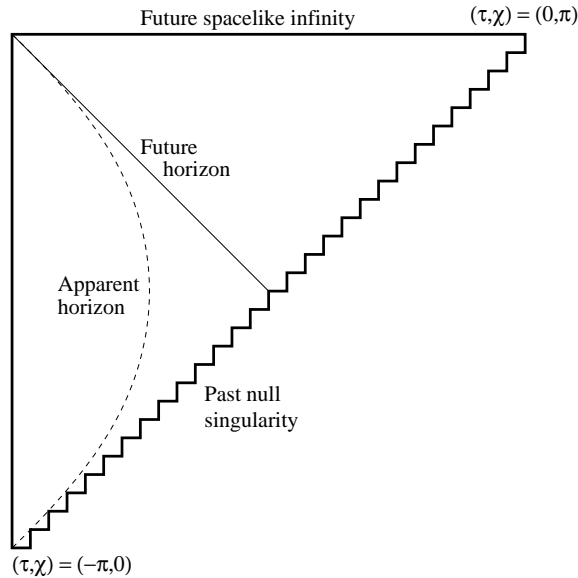}
\caption{Penrose diagram of the accelerating ($-1 < w < w\ll{\rm crit}$) universe.
The initial singularity is null, and the future spacelike boundary is obscured 
from observers by a horizon.}
\label{penrose1}
\end{center}
\end{figure}

In the accelerating case ($-1 < w < w\ll{\rm crit}$), the
singular hypersurface at $t = 0$ is
null, in the sense that it is diagonal on a Penrose diagram.
Future infinity is spacelike in this range, and all but one point of
the future boundary of the Penrose diagram is obscured
from the view of any observer by a horizon \cite{hks}.  
In particular, a horizon exists 
at a proper distance
\bbb
L\ll H = t {{(D - 1)(1 + w)}\over{\lba (D - 1) w + (D - 3) \rba}}\ .
\eee
This horizon recedes at a fixed proper speed, which increases to the speed of light 
as $w$ approaches $w\ll{\rm crit}$ from
below.  This is depicted in Fig.~\ref{penrose1}.  
(There is no horizon for $w \geq w\ll{\rm crit}$.)

\begin{figure}[htb]
\begin{center}
\includegraphics[width=1.7in,height=3.5in,angle=0]{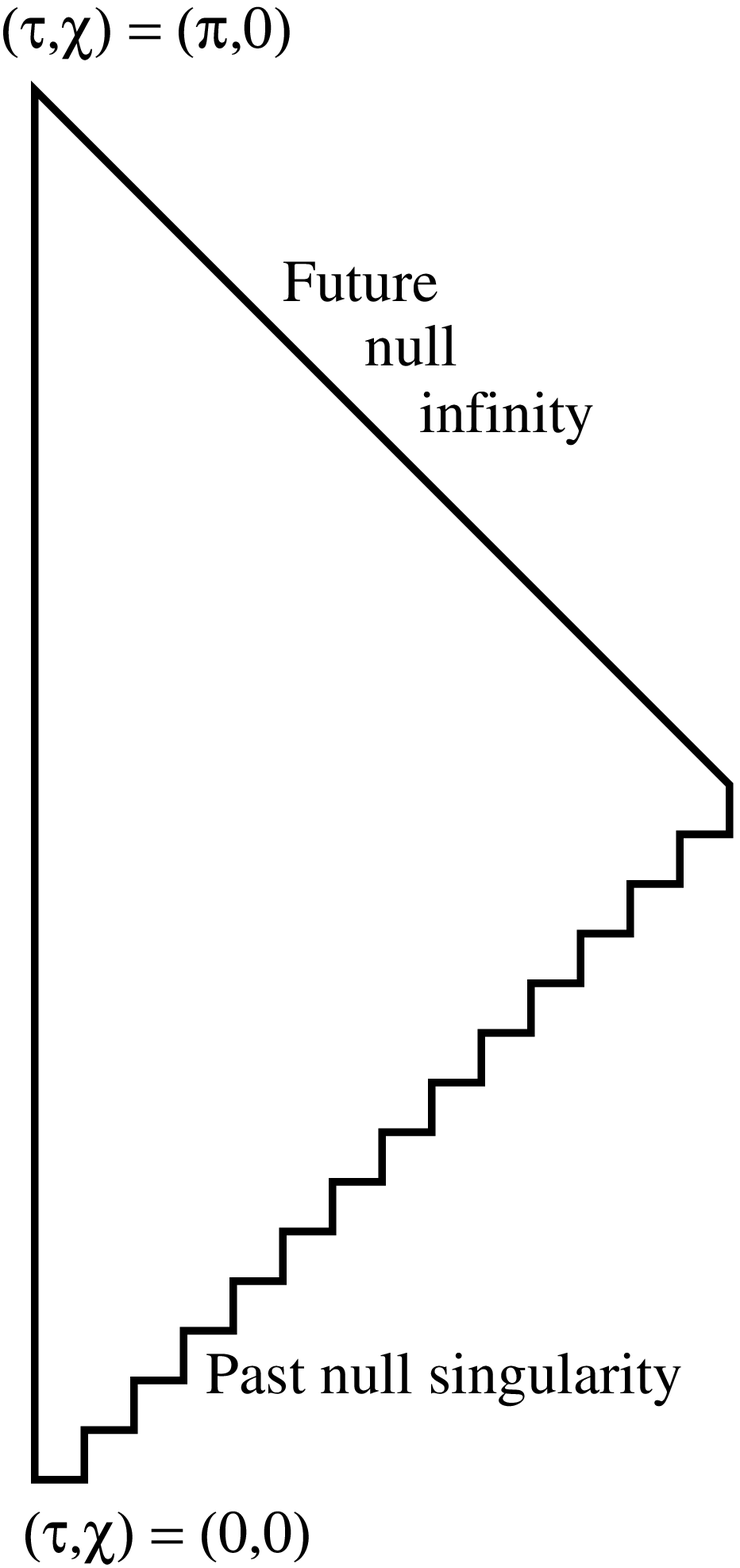}
\caption{Penrose diagram of the universe with critical equation of state ($w = w\ll{\rm crit}$).
The initial singularity is null, as is the future boundary. }
\label{penrose3}
\end{center}
\end{figure}

The liminal case $w = w\ll{\rm crit}$ is a hybrid of the two, with
a null initial singularity and a null future infinity; the Penrose 
diagram for this configuration is depicted in Fig.~\ref{penrose3}.  
In fact, the
liminal quintessent solution is globally conformally equivalent to 
Minkowski space, and we will find a simple way to understand this fact in our
stringy models.

\subsection{Spacetime effective
action for supercritical string theory}
We now introduce the spacetime effective action for string theory
formulated in a supercritical number of spacetime dimensions.  
Although we restrict our attention to the bosonic string, 
most of the discussion applies to the various closed superstring theories as well.
The exact effective action of any string theory is typically a complicated
expression involving nonrenormalizable interactions of arbitrarily high
dimension.  Much of the qualitative physics of string theory, however,
is captured by the dynamics of the effective action for light
fields, truncated to terms with few derivatives.

For the bosonic string in $D > 26$, the effective action for
the metric and dilaton appears as
\bbb
S_{\rm eff} = {1\over{2\k\sqd}} \int d\uu D x \sqrt{- \det G\us}
\exp{- 2\Phi} \lsqq - {{D - 26}\over{3\apr}} + R\us
+ 4 (\pp\Phi)\sqd \rsqq\ .
\een{effact}
Higher dimension terms are dropped:  
such terms in the tree-level action are suppressed by powers 
of $\apr = 1/ (2\pi T\ll{\rm string})$, where
$T\ll{\rm string}$ is the fundamental string tension.
Note that the action in Eqn.~(\ref{effact})
is written in terms of the string-frame metric $G\ll{\m\n}\us$.
We may rewrite the action in terms of the Einstein metric using the field redefinition in 
Eqn.~\rr{weyl}.  Using the rescaling
$\Phi \to \hh \sqrt{D - 2} ~\phi$,
the scalar field appears canonically, and we obtain the following form:
\bbb
S_{\rm eff} = {1\over{2\k\sqd}} \int d\uu D X \sqrt{- \det G\ue}
\left[ 
- {{2 (D - 26)}\over {3\apr}} ~\exp{{{2\phi}\over{ \sqrt{D - 2}}}}
+ R\ue - (\pp\phi)\sqd 
\right]\ .
\eee
We note that this is the action for a quintessent 
cosmology, taking the same form as Eqns.~(\ref{quaction}, \ref{einac}) above,
with coefficients now defined by the following values:
\bbb
\g = {2\over{\sqrt{D - 2}}}\ ,
\qquad
c = {{D - 26}\over{3 \apr}}\ .
\eee
We therefore recover a quintessent solution with equation of state given by
the formula in Eqn.~\rr{wformula}:
\bbb
w = - {{D - 3}\over{D - 1}}\ .
\eee
The tree-level potential of the string
theory thus gives rise to an equation of state at the boundary 
between accelerating and decelerating cosmological backgrounds. 
As we have seen above, the resulting spacetime is globally conformally 
equivalent to Minkowski space.

In retrospect, it is inevitable that this should be the case.  
The worldsheet theory of
the string in the timelike linear dilaton
background is {\it defined } to have a target space with string frame metric $\eta\ll{\m\n}$,
and coordinates $X\uu\m$ that are infinite in extent.  A spacetime potential that
encodes the worldsheet physics consistently must therefore produce a solution for which the metric
has those features.  The equation of state $w = - {{D - 3}\over {D - 1}}$ is the unique
value allowing such a solution.

One important point is that $w$ is negative for any $D\geq 4$. 
The matter and radiation equations of state are always 
$w_{\rm mat}=0$ and $w_{\rm rad} = + 1 / (D - 1)$,
respectively.  Both components have nonnegative $w$, but the component of the
stress tensor with the most negative value of $w$ always dominates the evolution of the 
universe at large scale factor.  
This means that particle production at early times never changes the late time behavior of
the solution.

The general quintessent FRW solution to this system is
\bbb
\phi &=& \phi\ll 0  -\sqrt{D - 2} \ln \lrdd { t \over{t\ll 0}} \rrdd\ ,
\nn\\ && \nn\\
a &=& {{a\ll 0}\over {t\ll 0}} t\ .
\label{atformula}
\eee
In terms of the original dilaton field $\Phi$, this solution takes the following form:
\bbb
\Phi = \Phi\ll 0 - {{D - 2}\over 2} \ln\lrdd { t \over{t\ll 0}} \rrdd\ .
\eee
Since the background metric is conformally flat, it is advantageous to convert to
a more natural time coordinate $t_{\rm conf}$ given by
\be
 t\ll{\rm FRW} &=& t\ll 0 ~
\exp{+ {{2 q ~t\ll{\rm conf} }\over{D - 2}}  }\ ,
\nn\\ \nn\\
t\ll{\rm conf} &=& + {{2 (D - 2) }\over q}
 \ln\left( \frac{t\ll{\rm FRW}}{ t\ll 0 }\right)\ ,
\ee
where, for convenience, we have defined a quantity $q$ according to
\bbb
 q \equiv  \sqrt{c\over 2} = 
\sqrt{ {{ (D - 26)}\over{6\apr }} }\ .
\eee
In these variables, the Einstein metric is manifestly conformally flat
and the dilaton is linear:
\be
ds\sqd &=& {{a\ll 0\sqd}\over{t\ll 0\sqd}}~t\sqd \left( - dt\ll
{\rm conf}\sqd
+ dx\uu i dx\uu i \right) = a\sqd \eta\ll{\m\n}dx\uu\m dx\uu\n\ ,
\\ \nn \\
\Phi &=& \Phi\ll 0 - q~t\ll{\rm conf}\ ,
\ee
with $x\uu 0 \equiv t\ll{\rm conf}$.

At this point we can set $a\ll 0$ and $t\ll 0$ at our convenience.
It turns out that a useful choice is for these quantities to be such
that the string-frame metric is unit-normalized.
Applying the Weyl transformation in Eqn.~\rr{weyl}, the string-frame
metric takes the general form
\bbb
G\ll{\m\n}\us = a\ll 0 \sqd ~\exp{{{4\Phi\ll 0}\over{D - 2}}} \eta\ll{\m\n}\ .
\eee
The condition $G\ll{\m\n}\us = \eta\ll{\m\n}$ therefore 
implies the following value for $a\ll 0$:
\bbb
a\ll 0 &=& \exp{ - {{2\Phi\ll 0}\over{D - 2}}}\ .
\eee
In turn, by demanding that $a^2 dt_{\rm conf}^2 = dt_{\rm FRW}^2$,
we find 
\be
a^2 = \frac{4 q^2}{(D-2)^2}\, t_{\rm FRW}^2\ .
\ee
From the second line in Eqn.~\rr{atformula} ($ a = a_0 t/t_0$), this translates into
the following value for $t_0$:
\bbb
t\ll 0 &=& {{(D - 2) a\ll 0 }\over{\sqrt{2 c}}}\ .
\eee

In conformal coordinates, the string-frame metric is trivial, and the dilaton
is linear in the timelike direction, with gradient
\bbb
\pp\ll{x\uu 0}\Phi = - q \equiv - \sqrt{{(D - 26)}\over{6\apr}}\ .
\een{dilgrad}
This agrees with the known formula for the strength of the dilaton
gradient in noncritical bosonic string theory (see, e.g.,~\cite{joebook}).

\subsection{Fluctuations of the metric and dilaton}
In time-dependent backgrounds there
is no obvious and natural definition of stability.
One definition would be to 
rescale modes in such a way that the kinetic terms 
are described by a trivial, canonically normalized kinetic term. 
One may then 
define an unstable mode as
one that grows exponentially in time (once
it has been rescaled to be canonical).
According to this definition, we will see that there indeed exist
`unstable' modes of the 
metric and dilaton in timelike linear dilaton theories
\cite{seiberg}.

However, the criterion of stability in terms
of the growth of canonical modes
is not completely satisfactory.  According to this criterion,
a spacetime-independent shift of a massless scalar 
(such as the dilaton) is counted as an unstable mode.  
Furthermore, various modes of the theory that are pure 
gauge would also be termed unstable,
such as the mode that implements
an overall rescaling 
of the metric, or the longitudinal modes of the NS/NS B-field.

By taking into account the coupling of the background 
fields to modes of the string, we can formulate a definition
of stability that avoids these difficulties \cite{ali}.
The coupling of a background field to the
string is 
suppressed by a factor of $g\ll s = \exp{\Phi}$.  If
an exponentially growing mode grows more slowly
than $g\ll s\uu{-1}$, its effect on the remaining degrees of
freedom in the theory therefore decreases in time.  
We may thus define an unstable (stable) mode 
as one that grows faster (slower) than $g\ll s\uu{-1}$ at late times.
This indeed alleviates the problems noted above: we can see that longitudinal modes
of gauge fields have 
\bbb
({\rm amplitude}) / g\ll s \sim {\rm const.}
\eee
at $t\to + \infty$, and the same holds for the mode representing a constant shift
of the dilaton $\Phi \to \Phi + {\rm const.}$

To illustrate this point more explicitly,
let us work out the equations of motion for a
massless scalar $\sigma$ coupled with unit strength to the 
fundamental string:
\be
{\cal L}_\sigma = -\frac{1}{2 \kappa^2} \sqrt{-\det G\us} e^{-2\Phi} (\del \sigma)^2
	= -\frac{1}{2\kappa^2}\sqrt{- \det G^{(E)}} (\pp\s )^2\ .
\ee
The scalar field has solutions of the form
\be
\sigma = \sigma_\infty - \xi \, t^{-(D-2)}\ ,
\label{sigmanew}
\ee
where $\s\ll{\infty}$ and $\xi$ are constants of motion that
can take arbitrary real values.
Eqn.~\rr{sigmanew} states that the modes of the field $\sigma$ asymptote to 
the constant value $\s\ll{\infty}$ as $t\to \infty$.
From the point of view of the Einstein frame, this effect is due to Hubble friction;
in the string frame this behavior is understood to be caused by the drag force arising from the 
interaction between $\s$ and the linear dilaton.

As it stands,
$\sigma $ is coupled to an Einstein metric that is conformally flat,
albeit with a nontrivial conformal factor.
A state of the string is necessarily coupled to the flat metric, as 
the kinetic term of the string sigma model is the flat metric $G^{(S)}$.
To recover field 
fluctuations whose quanta represent normalizable string states,
we must introduce a rescaled field $\tilde \sigma$:
\be
\tilde \sigma \equiv e^{-\Phi }\sigma \ .
\ee  
This induces a mass term for the rescaled field
that couples to a trivial metric and represents a proper quantum of string.
Schematically, we have  
\be
e^{-2\Phi}(\del\sigma)^2 = (\del\tilde\sigma)^2 + 
\tilde\sigma^2(\del\Phi)^2
+  2\tilde\sigma (\del\tilde\sigma)
\lsqq (\del\Phi)_{\rm background} + (\del\Phi)_{\rm fluctuation} 
\rsqq
	\ .
\ee
The fluctuation term represents a trilinear vertex that we discard from the outset.
The background term is constant, so its
dot product with $\tilde{\s}\, \pp \tilde{\s}$
amounts to a total derivative.
The mass term for the rescaled field is tachyonic 
and proportional to $-q\sqd$, and the quadratic action for $\tilde{\s}$ thus reads
\be
{\cal L}_{\tilde{\s} } \sim -\frac{1}{2\kappa^2}\left[
	(\del\tilde\sigma)^2 - q\sqd \tilde\sigma^2\right]\ .
\ee

The criterion for stability we proposed was that
$g\ll s$ times the canonical field $\tilde{\s}$ not
increase exponentially with time.  Since $g\ll s \tilde{\s}$
is just $\s$, the original field appearing in the
spacetime action in front of $\exp{- 2 \Phi}$, the requirement
for physical stability is that 
modes normalized to have the factor $\exp{-2\Phi}$ in their kinetic
term should shrink exponentially in the future or, at most,
remain constant in magnitude.

At this stage, a simple exercise is to find
linearized modes of $\tilde{\s}$ and $\s$ that are plane
waves in the spatial directions:
\bbb
\tilde\sigma = {\cal A}\sin(\vec{k}\cdot\vec{x})   
e^{\pm \tilde{\Gamma}\, t_{\rm conf}}\ ,
\een{first}
or
\be
\tilde\sigma = {\cal A} \sin(\vec{k}\cdot\vec{x})   e^{i\, \omega\, t_{\rm conf}}\ ,
\ee
where
\be
\tilde\Gamma^2  =  q\sqd - \vec{k}^2\ ,
\qquad
\o\sqd  =  \vec{k}\sqd - q\sqd\ ,
\ee
and ${\cal A}$ is an arbitrary mode amplitude.
Moving back to FRW time and the $\sigma$ normalization for the scalar field, we recover the form
\be
\sigma = e^{\Phi_0}{\cal A}\sin(\vec{k}\cdot\vec{x})   \left(\frac{t}{t_0}\right)^{\cal B\ll\pm}\ ,
\ee
for the overdamped modes,\footnote{These modes are referred to as ``pseudotachyons''
in \cite{evaofer}.} where
\be
{\cal B}\ll\pm & \equiv &
  {{(D-2)}\over{2 q}} \Gamma_\pm\ ,
\nn\\ \nn\\
\Gamma_\pm & \equiv &  \pm 
\sqrt{q\sqd - \vec{k}^2} - q\ .
\label{last}
\ee

At $k = 0$ and $\Gamma_+ = 0 $, a background value of
the mode 
represents a ``condensation'' of the massless field, 
which shifts the asymptotic value $\sigma_\infty$.
At the $k = 0$, $\Gamma_- = -2 q$ end, we have
\be
\sigma = e^{\Phi_0}\left( \frac{t}{t_0}\right)^{-(D-2)}\ .
\ee
As such, the underdamped modes of the massless
field are given by
\bbb
\s = \exp{\Phi\ll 0} {\cal A} \sin{\vec k \cdot \vec x}
~\lrdd {{t \over {t\ll0}}} \rrdd\uu{-{{D - 2}\over 2}}
~\lrdd {t\over{t\ll 0}} \rrdd \uu{{\pm {{i (D - 2)\o}\over {2q}}}}\ .
\eee
It is clear that these are all stable, in the sense
that they asymptote to zero at late times.

Upon adding a mass term of the form
\bbb
-\hh \exp{- 2\Phi} \sqrt{-\det G\uu{(S)}} ~m\sqd \s\sqd
\nn
\eee 
to the Lagrangian,
we find that the system again breaks into a set
of overdamped modes with $\vec{k}\sqd < q\sqd - m\sqd$,
and underdamped modes with $\vec{k}\sqd > q\sqd + m\sqd$.
Under the replacement
$\vec{k}\sqd \to \vec{k}\sqd + m\sqd$, the form of these modes is
the same as those in Eqns.~(\ref{first}---\ref{last}),
and all $\s$-modes decay exponentially in time when $m\sqd \geq 0$.

When a mode acquires a background value obeying the equations of motion, 
we may ask what the
resulting effect is on string propagation as $t\to \infty$.   
The specific effect from the background is to add
a term 
\bbb
{\cal O}\, \exp{ \Phi(t)} \tilde{\s}(x,t)
= \s (x,t)
\nn
\eee 
to the string worldsheet action, where ${\cal O}$ is
some operator in the 2D conformal field theory describing the dynamics of a
string.
In all cases, $\s$ decreases exponentially or
stays constant as long as $m^2 \geq 0$.  The effect of the mode on
string propagation decreases exponentially with time, 
or at most stays constant as $t\to\infty$.
It is therefore best to not think of these modes as describing physical instabilities
of any kind, despite the fact that the quantum mechanics problem for the
canonical mode $\tilde{\s}$ formally describes a field with exponential growth in time.  
The overdamped modes, for which the canonically normalized field $\tilde{\s}$ has
exponentially growing modes, correspond to {\it non-normalizable } states
of the string.  These modes do not have an interpretation as
particle excitations of string theory.  

Only when the
$\s$ field is a {\it tachyon}, in the sense that its
quadratic term in the action is
$$
-\hh \exp{- 2\Phi} 
\sqrt{-\det G\uu{(S)}} ~m\sqd \s\sqd\ ,
$$
with $m\sqd < 0$, does the $\s$ field have
an exponentially growing mode.
If $\vec{k}\sqd  < |m\sqd|$, we have solutions of the form
\bbb
\s &=& \exp{\Phi\ll 0} {\cal A} \sin{\vec{k} \cdot \vec{x}}
\lrdd {t\over{t\ll 0}} \rrdd \uu{{\cal B}\ll \pm}\ ,
\nn\\ && \nn\\
{\cal B}\ll\pm &=& {{D - 2}\over{2 q}} \G\ll\pm\ ,
\nn\\ && \nn\\
\G\ll \pm &=& - q \pm \sqrt{q\sqd + |m\sqd| - \vec{k}\sqd }\ .
\label{tacheoms}
\eee
To be specific, the field in the first line of Eqn.~\rr{tacheoms}, 
corresponding to ${\cal B}\ll +$, grows exponentially with time. 
In a generic
number of dimensions, most anomaly-free
string theories have tachyons of this kind.
The bosonic string has a tachyon ${\cal T}$ whose mass $m$ satisfies
$m\sqd = - {4\over\apr}$.
In the next section we will consider particular solutions for which
the bosonic string tachyon acquires an expectation value.

\section{Worldsheet description of supercritical string cosmology}
As pointed out in the introduction, string theory in $D> D\ll{\rm crit}$ is 
most easily studied in
the timelike linear dilaton background.\footnote{In $D\neq D\ll{\rm crit}$, objects such
as branes or fluxes can create a stationary point in the dilaton potential, a
possibility that was studied in \cite{Silverstein:2001xn,Maloney:2002rr}.  
Such theories are not conveniently
described by a worldsheet conformal field theory,
and their $\apr$ corrections are not entirely understood.}
For the sake of clarity and convenience, and to introduce our analysis
in the language of worldsheet conformal field theory, we briefly record a 
few basic and well-known facts about this theory.

The worldsheet description in conformal gauge is a
two-dimensional theory of $D$ free, massless scalars $X\uu\m$ playing the role of the 
embedding coordinates of spacetime.  The linear dilaton enters as a modification of
the definition of the two-dimensional stress tensor, relative to the $SO(D-1,1)$-invariant
definition.  That is, for
a dilaton with gradient $\pp\ll \m\Phi = V\ll\m = {\rm const.}$,
the two-dimensional stress tensor is 
%
%
\be
T_{++} &=& - {1\over{\apr}} :\pp\ll{\s\uu +} X\uu\m \pp\ll{\s\uu +} X\ll\m :
+ \pp\sqd\ll{\s\uu +} (V\ll\m X\uu\m)\ ,
\nn\\ \nn\\
T_{--} &=& - {1\over{\apr}} :\pp\ll{\s\uu -} X\uu\m \pp\ll{\s\uu -} X\ll\m :
+ \pp\sqd\ll{\s\uu -} (V\ll\m X\uu\m)\ ,
\ee
where the colons represent normal ordering of the $2D$ theory.
Here, $\s\uu\pm$ are particular light-cone combinations of the worldsheet
coordinates $\s\uu{0,1}$:
\bbb
\s\uu\pm = - \s\uu 0 \pm \s\uu 1\ .
\eee

This worldsheet theory completely defines the dynamics of a string
propagating in the background of a timelike linear dilaton.  
Physical states of the string correspond to
local operators ${\vertexoperator}$ that are Virasoro primaries of weight one.
That is, their operator product expansion (OPE) with the stress tensor satisfies:
\bbb
T\ll{++}(\s) {\vertexoperator}(\t) \simeq
{{{\vertexoperator}(\t)}\over{{(\s\uu + - \t\uu +)\sqd}}}
+ {{{\pp\ll + {\vertexoperator}(\t)}}\over{\s\uu + - \t\uu +}}\ ,
\eee  
and similarly for $T\ll{--}$, 
where $\simeq$ denotes equality up to terms that are
smooth as $\s\to\t$.

%
%

The linear dilaton theory is free, and it exists as a CFT to all orders, and
nonperturbatively, in $\apr$.  The theory represents a solution to the equations of motion
at leading order in derivatives.  Since the solution is exact, however, it must 
satisfy the equations of motion to all orders in derivatives.  
As higher derivative corrections to the string effective action are not 
known in closed form, the existence of such a solution constrains the form of 
any higher derivative corrections to the leading effective action.
In this section we consider a novel set of exact solutions that deform the linear dilaton
background and exhibit a nonvanishing tachyon.  Following the example of the strict linear dilaton theory,
we will use the existence (and exactness) of our new solutions to constrain
the effective action, to leading order in derivatives and to all orders in the tachyon field ${\cal T}$.


\subsection{Exact solutions with nonzero tachyon}
We aim to study solutions 
for which the bosonic string tachyon is not everywhere equal to zero.  
To linearized order in the strength of the tachyon field,
the deformation of the background is described in the worldsheet theory
by the insertion of a single tachyon vertex operator into all
correlation functions.  The spatial component of the vertex operator is determined by the
spatial dependence of the background deformation; the condition
that the linearized deformation of the background satisfy the equations of 
motion is simply the condition that the tachyon momentum be on-shell.
This, in turn, equates to the condition that the matter part of the
vertex operator ${\vertexoperator}\ll M$, made of the $X\uu\m$ degrees of freedom, 
be a conformal primary of weight $(1,1)$.\footnote{We will omit discussion of the 
ghost contribution to the vertex operator.}

To be specific, a profile ${\cal T} (X)$ for
the tachyon corresponds to the vertex operator
\bbb
{\vertexoperator}\ll M \equiv :{\cal T}(X):\ ,
\eee
and admits the following on-shell condition: 
\bbb
\pp\ll\m \pp\uu\m {\cal T}(X) - 2 V\uu\m \pp\ll\m {\cal T}(X) + {4\over{\apr}} {\cal T}(X) = 0\ ,
\een{teom}
where $V\ll\m$ is defined, per convention, as the dilaton gradient.  
For tachyon profiles of the
form ${\cal T}(X) = \m\sqd \exp{B\ll\m X\uu\m}$, this condition is
$B\sqd - 2 V\cdot B = - 4/\apr$.   A general value of $B\ll\m$ will lead
to a nontrivial interacting theory when the strength $\m\sqd$ of
the perturbation is treated as non-infinitesimal.  For instance, the insertion
of two copies of the vertex operator can lead to singularities when the
positions of the vertex operators approach one another;  counterterms may
have to be added to preserve conformal invariance, thus complicating the
analysis of the theory considerably.

There is a special set of choices for $B\ll\m$ satisfying the equations
of motion such that the $2D$ theory is well-defined and conformal to all orders in 
perturbation theory, as well as nonperturbatively, in $\apr$.   
In particular, we can choose the first term in the linearized
tachyon equation of motion in Eqn.~\rr{teom} to vanish separately, which is 
tantamount to choosing the vector $B\ll\m$ to be null.
It is easy to see that this renders the vertex operator $:\exp{B\ll\m X\uu\m}:$ 
non-singular in the vicinity of itself.  In a free field theory, all
singularities of normal-ordered operators arise from propagators contracting
a free field in one operator with a free field in the second.  If both
operators depend only on $B\cdot X$, however, all contractions are 
proportional to $B\ll\m B\uu\m$, which vanishes when $B_\mu$ is null. 

We therefore employ a Lorentz transformation to put $B\ll\m$ into the form
\be
B\ll 0 &=& B\ll 1\ \equiv \ {\b/{\sqrt{2}}}\ ,
\\ \nn \\
B\ll i &=& 0, \qquad i\geq 2\ .
\ee
Letting
\bbb
X\uu\pm \equiv {1\over{\sqrt{2}}} (X\uu 0 \pm X\uu 1)\ ,
\eee
the vertex operator becomes ${\vertexoperator}\ll M  = \m\sqd \exp{\b
X\uu +}$.  At this point we can drop the normal-ordering symbol because
self-contractions of the null fields in the exponential
are zero.

\begin{figure}[htb]
\begin{center}
\includegraphics[width=1.5in,height=1.0in,angle=0]{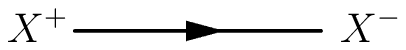}
\caption{Oriented propagators are drawn with a directed line
between $X\uu +$ and $X\uu -$.}
\label{feyn1}
\end{center}
\end{figure}

To deform the background by a finite amount, we
add ${\vertexoperator}\ll M$ to the worldsheet Lagrangian:
this gives rise to a particularly simple quantum theory.
The kinetic term for $X\uu\pm$ appears as 
\bbb
{\cal L} \sim 
- {1\over{2\pi\apr}} 
\lsqq
(\pp\ll {\s\uu 0} X\uu +)(\pp\ll {\s\uu 0 }X\uu -)
- (\pp\ll {\s\uu 1} X\uu +)(\pp\ll {\s\uu 1} X\uu -) \rsqq\ .
\eee
The propagator for the $X\uu\pm$ fields is therefore oriented, in the
sense that every propagator has an $X\uu +$ at one end and an $X\uu -$
at the other end.  A propagator can thus be depicted by a
line with an arrow pointing from $X\uu +$ to $X\uu -$, as depicted in
Fig.~\ref{feyn1}.
The tachyon couples to the worldsheet according to
\bbb
{\cal L} \sim - {1\over{2\pi}} \m\sqd \exp{\b X\uu +}\ , 
\eee
and the equations of motion for the string appear as
\bbb
\pp\ll + \pp\ll - X\uu i &=& \pp\ll - \pp\ll + X\uu + = 0, ~~~~~i = 2,3,\cdots,D - 1\ ,
\nn \\ && \nn \\
\pp\ll + \pp\ll - X\uu - &=& + {{\apr \b M\sqd}\over 4}\ ,
\eee
where $M\sqd \equiv \m\sqd \exp{\b X\uu +}$.

\begin{figure}[htb]
\ \ \ \ \ \ \ \includegraphics[width=4.4in,height=1.0in,angle=0]{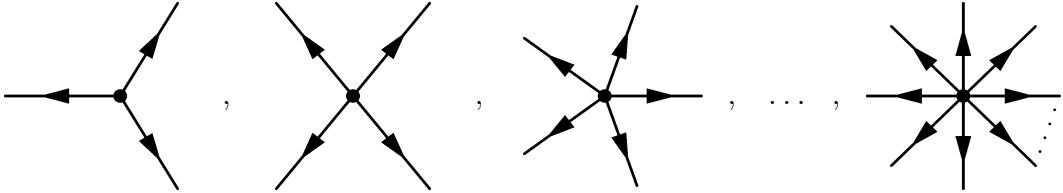}
\caption{All interaction vertices correspond to diagrams composed of 
outgoing lines.}
\label{feyn2}
\end{figure}

By writing the solution to the Laplace equation for $X\uu +$ as 
\bbb
X\uu + = f\ll +(\s\uu +) + f\ll - (\s\uu -)\ ,
\eee  
for arbitrary $f_\pm$,
the general solution for $X\uu -$ can be expressed as follows:
\bbb
X\uu - = g\ll + (\s\uu +) + g\ll - (\s\uu -) + {{\apr \b \m\sqd}\over 4}
\left[ \int\ll {\s\uu +}\uu\infty dy\uu + \exp{\b f\ll +(y\uu +)} \right] 
\left[ \int\ll {\s\uu -}\uu\infty   dy\uu - \exp{\b f\ll -(y\uu -)} \right]\ ,
\nn\\
\eee
where $g_\pm(\sigma^\pm)$ are arbitrary functions.
We thus see that the theory is exactly solvable at the classical 
level.  It is clear, however, that the classical solvability extends 
neatly to the quantum regime.  All interaction vertices in the theory
depend only on $X\uu +$, and therefore correspond to diagrams
composed strictly from outgoing lines (see Fig.~\ref{feyn2}).
Nontrivial Feynman diagrams, including tree graphs, must have at least two
vertices connected by an internal line.  Since all such diagrams are absent, 
we obtain a fully quantum-mechanical theory in which all amplitudes are given by 
their classical limits, with no quantum corrections of any kind. 
Note that this is only possible in a nonunitary theory:  
in a unitary quantum theory, sewing rules dictate nonzero values for tree and 
loop graphs, given a nonvanishing set of interaction 
vertices.\footnote{Solutions similar to the type studied here were discussed previously
in various contexts \cite{Tseytlin:1992pq,Tseytlin:1992ee,Russo:1992yh,Callan:1992rs,Russo:1992yg,Verlinde:1991rf,Burwick:1992yx,Strominger:1992zf,deAlwis:1992as,Bilal:1992kv,Giddings:1992ae}.
We thank A.~Tseytlin for bringing these to our attention. }

\begin{figure}[htb]
\begin{center}
\includegraphics[width=4.0in,height=1.1in,angle=0]{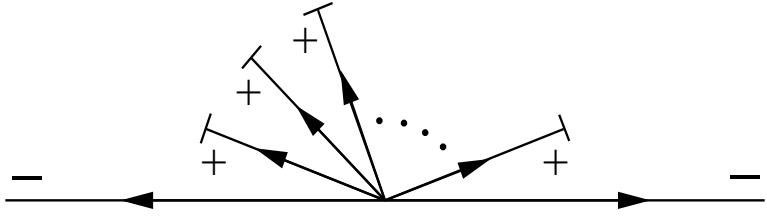}
\caption{Generic structure of OPEs in the lightlike tachyon background.}
\label{ope}
\end{center}
\end{figure}

The structure of the OPE
is correspondingly simple.  By virtue of the
Feynman rules, all interaction contributions to
OPEs are determined by
those involving $X\uu -$ alone.  That is to say,
the OPE of $X\uu +$ with any other local operator
is unmodified. Likewise,
interaction corrections to OPEs of $X\uu -$
have only operators involving $X\uu +$ on the right-hand side,
of the form depicted in Fig.~\ref{ope}.
For example, we have the following relations:
\bbb
X\uu +(\s) X\uu + (\t) &=& ~:X\uu +(\s) X\uu + (\t) :\ ,
\nonumber
\\
\nonumber
\\
X\uu + (\s) X\uu - (\t) &=& ~: X\uu + (\s) X\uu - (\t) :
+ {{\apr}\over 2} \ln \lba - {1\over{L\sqd}}
(\s\uu + - \t\uu +)(\s\uu - - \t\uu -) \rba \ ,
\nonumber
\\
\nonumber
\\
X\uu -(\s) X\uu - (\t)
&=& ~: X\uu -(\s) X\uu - (\t)  :
\nonumber
\\
&&
\nn\\
&&
\kern-40pt
+~{{\apr\b\sqd}\over 8} \ln \lba - {1\over{L\sqd}}
(\s\uu + - \t\uu +)(\s\uu - - \t\uu -) \rba 
\cdot \int\ll{\s\uu +}\uu{\t\uu +}
dy\uu + \int\ll{\s\uu -}\uu{\t\uu -} dy\uu -
 \m\sqd \exp{\b X\uu +} 
\nn\\
&&
\nn\\
&&
\kern-40pt
+~({\rm terms~that~remain~smooth~as}~\s\to\t)\ .
\eee

We note the similarity of our theories to those 
of strings propagating on plane waves
\cite{horsteif1,horsteif2,hellfab},
in which the moduli of string theory vary along lightlike
directions.  The structure of Feynman diagrams,
and the controlled nature of the quantum corrections
are similar.  In fact, the theories of \cite{horsteif1,horsteif2} are
solvable for precisely the same reasons as the theories in this
paper: the interaction vertices depend on only one
of the light-cone directions, and therefore have only
outgoing lines.  The interaction vertices thus
have no Wick contractions
with themselves or each other through lines associated with the
light-cone coordinates.  The primary difference is
that the interaction vertices in our model represent 
a {\it massive} (yet conformal) 
perturbation of the $2D$ CFT, rather than a massless one.



\subsection{Physical interpretation of the solution}
The solution ${\cal T}(X) = \m\sqd \exp{\b X\uu +}$
can be thought of as
a phase boundary in spacetime between a ${\cal T} \sim 0$ phase
and a ${\cal T} > 0$ phase.  The boundary is
moving to the left at the speed of light.
Intuitively, the ${\cal T}> 0$ phase is a
`nothing' phase, into which no particles,
including the graviton itself, can enter.  
One way to interpret this phase is 
to view it as an alternate description of
the absence of spacetime itself.  

A related
spacetime-destroying solution
is given by the
Lorentzian continuation of the
gravitational instanton of \cite{wittenbubble}.
The physics of this solution can be described by 
first noting that the Euclidean solution exhibits time reversal symmetry
at one point, say $t = 0$.
At that point, one matches the
Euclidean instanton onto a Lorentzian
solution.  After that point in Lorentzian time,
a bubble of nothing expands, rapidly accelerating 
to the speed of light at constant proper acceleration.

Viewed from the outside,
the physics is very much the same as that of an
ordinary Coleman-de Luccia instanton, which mediates
tunneling between vacua, except that on the
inside of the bubble there are no degrees of freedom at all.
As a result, all matter that encounters the bubble wall
is rapidly swept away and accelerated to the speed
of light, along with the wall itself.
It has been conjectured that 
the solution of \cite{wittenbubble}
has an alternate description in terms of
closed-string tachyon condensation, and evidence for this
connection has been studied in various
settings \cite{Horowitz:2005vp,Hirano:2005sg,Headrick:2004hz,Gutperle:2002bp,Emparan:2001gm,Costa:2000nw,Adams:2005rb,Aharony:2002cx,zamolunpub,Horowitz:2006mr,McGreevy:2005ci}.

\begin{figure}[htb]
\begin{center}
\includegraphics[width=3.2in,height=3.2in,angle=0]{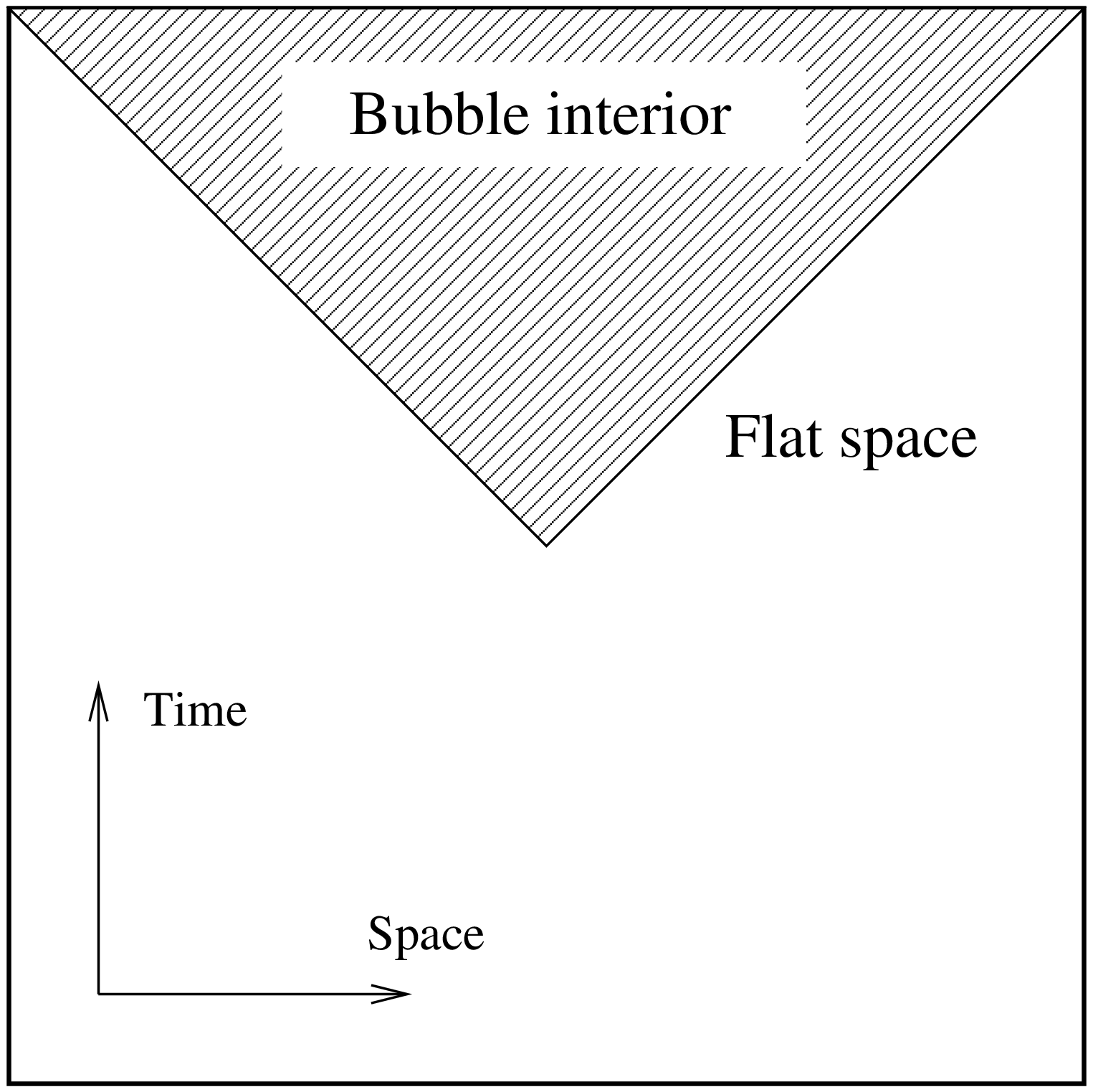}
\caption{A bubble of nothing, which destroys spacetime.
If the exterior is flat with vanishing dilaton gradient,
the bubble properly accelerates all the way to future infinity.
If the exterior has a flat string frame metric and timelike
linear dilaton gradient, the spatial thickness of the diagonal line
approaches $\apr | \dot{\Phi} | $ as $t\to\infty$.  Our solution corresponds
to a limit that focuses on the upper left-hand corner of the diagram.}
\label{bubblediagram}
\end{center}
\end{figure}

A full, classical solution for the closed string tachyon
description of the bubble of nothing is not known 
explicitly. 
We propose that in the presence of a
timelike linear dilaton, such a solution
would approach the lightlike solution we describe in
this paper at times long after the nucleation of the bubble.
We should note, however, that our
solution behaves differently from the bubble
of nothing in flat space in one
important respect: the wall in our 
model is not properly accelerating.  This is
a result of Hubble friction and
the presence of the timelike linear dilaton  
(see Fig.~\ref{bubblediagram}).
The drag force of the background fields stops the acceleration when the
spatial thickness of the bubble is of order 
$\apr | \dot{\Phi} |  \sim  \b\uu{-1}$.  In the absence of
a dilaton gradient, the bubble wall continues to
accelerate, and its spatial thickness in the frame
of a given static observer blueshifts to zero as $t\uu {-1}$,
where $t$ is the time since nucleation of the bubble.





We would also like to establish that the tachyon condensate
is indeed a `nothing' state, in which no degrees of freedom
can live.  This is intuitively obvious, since the high potential
barrier would seem to suppress the wavefunctions of 
any string state that might try to penetrate the
interior of the bubble.  Since our theory is exactly solvable,
we can do better by solving directly for the trajectory of a string
colliding with the bubble wall.

Let us consider a simple example for which the
string is pointlike, wherein all embedding coordinates
are independent of $\s\uu 1$ and depend only
on $\s\uu 0$.  The solution is then characterized
by the conserved momenta ${\momp}\uu i \equiv {1\over\apr} \Xd\uu i$
and ${\momp}\uu + =  {1\over{\apr}} \Xd\uu +$.
For a pointlike solution, 
$X\uu + = \apr {\momp}\uu + (\s\uu 0 - \s\uu 0\ll 0) + X^+_0$,
and $X\uu -$ is given by
\bbb
X\uu - = \apr {\momp}\uu -\ll {\rm initial}(\s\uu 0 - \s\ll 0\uu 0 ) +
{{ \m\sqd }\over{\b\apr
 {\momp}\uu{+2}}} \exp{\apr \b {\momp}\uu +(\s\uu 0 - \s\ll 0\uu 0)}
	+ X^-_0\ ,
\eee
where $ {\momp}\uu -\ll {\rm initial}$ and $X^\pm_0$ are constants of motion.
Classically, the Virasoro constraint is
\bbb
H\ll{\rm worldsheet} = -
\apr
 {\momp}\uu - {\momp}\uu + +\hh\apr {\momp}\ll i\sqd + 
\m\sqd \exp{\b X\uu +} = 0\ ,  
\eee
which means that
\bbb
{\momp}\uu - \ll{\rm initial} = {{{\momp}\ll i \sqd}\over{2 {\momp}\uu +}}\ .
\eee

\begin{figure}[htb]
\begin{center}
\includegraphics[width=4.0in,height=2.4in,angle=0]{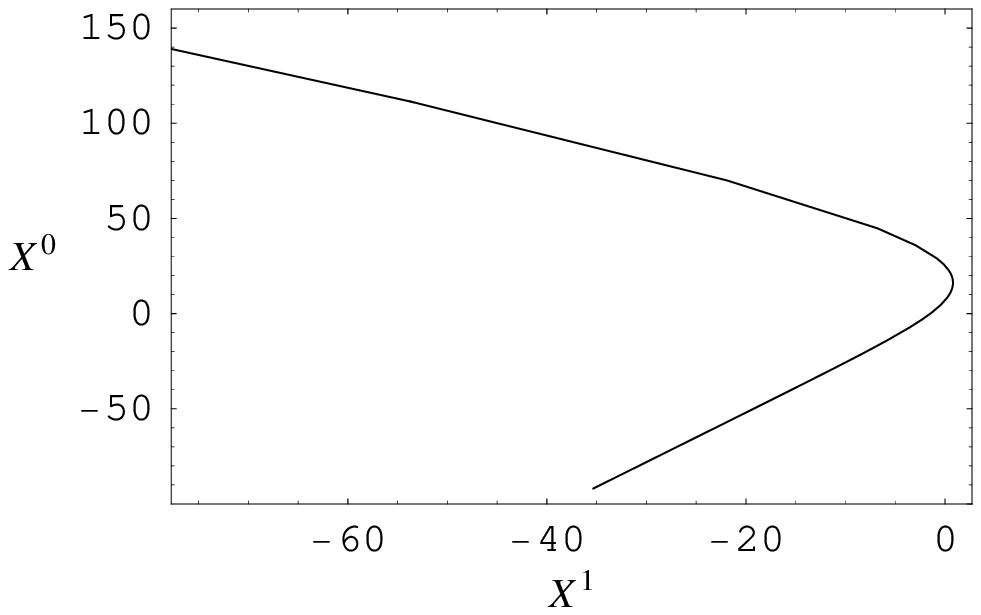}
\caption{A plot of the position of a particle pushed by the bubble wall, as a function
of time.  The solution has $\m\sqd = 1,~ \b = .1,~X^\pm_0 = 0$, and
the trajectory corresponds to $p\uu + = 3,~ H\ll{\perp} \equiv
{{\apr p\ll i\sqd}\over 2} = 4$.  We set $\apr = 1$.}
\label{positionplot}
\end{center}
\end{figure}

\begin{figure}[htb]
\begin{center}
\includegraphics[width=4.0in,height=2.4in,angle=0]{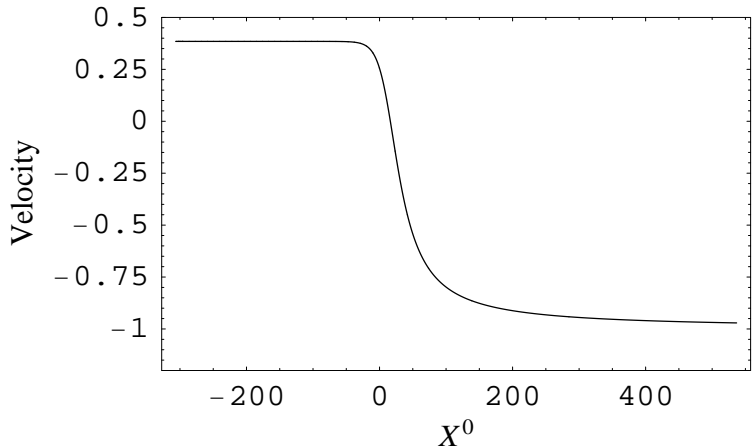}
\caption{A plot of the velocity of a particle accelerated by the bubble wall, 
as a function of time. (Numerical values are set in accordance with the
trajectory depicted in Fig.~\ref{positionplot} above.)}
\label{velocityplot}
\end{center}
\end{figure}

The physical interpretation of the solution is
clear:  the particle propagates 
at an initial speed 
\bbb
v\equiv {{\Xd\uu 1}\over{\Xd\uu 0}} 
= {{{\momp}\uu + - {\momp}\uu -\ll{\rm initial}}\over{{\momp}\uu + + {\momp}\uu - \ ,
\ll{\rm initial} }}
\eee
until it hits the bubble wall, 
where the exponential term becomes important.
At that point, the speed of the particle rapidly
goes to $-1$, as both the numerator and denominator
are dominated by the exponential term in $\momp\uu -$:
\bbb
\momp^- = \momp_{\rm initial}^- + \frac{\mu^2}{\alpha' \momp^+}\exp{ \alpha'\beta \momp^+(\s^0 - \s^0_0)}\ .
\eee
The position and velocity trajectories of such a particle are plotted in
Fig.~\ref{positionplot} and Fig.~\ref{velocityplot}, respectively.



\subsection{Effective actions for the dilaton-tachyon-metric system}
Effective actions for light fields in string theory,
truncated to operators with low numbers of derivatives,
often give a good qualitative description of some aspects
of string physics.  Although it is not entirely clear why
the truncation to a two-derivative effective action need
contain any useful information, the
leading action for the metric and dilaton correctly describes
many important aspects of the linear dilaton background,
even quantitatively.  Effective actions for this same
system, augmented by tachyons, fluxes, etc.,~have been 
successfully employed to describe closed-string tachyon physics
\cite{newhat,Minahan:2000tg,Minahan:2000tf,Minahan:2000ff,bergman,moller,freedman,takayanagi,Suyama:2002ky,Sen:2004nf,Tseytlin:2000mt,Yang:2005rx}. 
Furthermore, effective actions for the metric,
dilaton and tachyon have been used to describe the physics of 
dynamical dimension change in noncritical string theories
(see, e.g.,~\cite{bergman,freedman}).

Although these theories approximate the
normalization of the Einstein term as
a constant, independent of the tachyon condensate, it is
unclear how realistic this may be.  In theories of
open-string tachyon condensation, the analogous
dependence of the gauge field kinetic term on the
tachyon must be nontrivial to correctly 
reproduce the qualitative physics of kink solutions and of 
the closed-string vacuum.  We are therefore prompted to investigate 
constraints on the tachyon dependence of 
the Einstein action in a useful effective theory 
of closed-string tachyon condensation.

The definition of tree-level string amplitudes dictates that the
dilaton dependence of the tree-level action appear as
an overall factor equal to $\exp{- 2\Phi}$. 
The most general two-derivative ansatz for
the action with this dilaton dependence is then
\be
S = \frac{1}{2\kappa^2}\int d^D x \sqrt{\det G}\Bigl[
	{\cal F}_1 R - {\cal F}_2 (\nabla \Phi)^2 - {\cal F}_3 (\nabla {\cal T})^2
	- {\cal F}_4 - {\cal F}_5 \nabla {\cal T} \cdot \nabla \Phi \Bigr]\ ,
\ee
where the functions ${\cal F}\ll i$ are defined in 
terms of five arbitrary functions of the tachyon as follows:
\be
{\cal F}_1 & \equiv & e^{-2\Phi} f_1({\cal T})\ , \nn\\
{\cal F}_2 & \equiv & -4 e^{-2\Phi} f_2({\cal T})  \ , \nn\\
{\cal F}_3 & \equiv & e^{-2\Phi} f_3({\cal T})  \ , \nn\\
{\cal F}_4 & \equiv & 2 e^{-2\Phi} {\newv}({\cal T})  \ , \nn\\
{\cal F}_5 & \equiv & e^{-2\Phi} f_5({\cal T})  \ .
\ee
(For convenience, we have absorbed the $\exp{- 2\Phi}$ prefactor into 
the ${\cal F}\ll i$.)
The Einstein equation appears as
\be
0 & = & \left(\nabla^\mu \nabla^\nu - G^{\mu\nu} \nabla^2 
	+ \frac{1}{2} G^{\mu\nu} G^{\rho\sigma} R_{\rho\sigma}
	- R^{\mu\nu}\right) {\cal F}_1
	- \frac{1}{2}G^{\mu\nu}{\cal F}_4
	+ {\cal F}_2\nabla^\mu \Phi \nabla^\nu \Phi
\nn\\
&&	- \frac{1}{2}G^{\mu\nu} {\cal F}_2 (\nabla \Phi)^2
	+ {\cal F}_3 \nabla^\mu {\cal T} \nabla^\nu {\cal T}
	- \frac{1}{2} G^{\mu\nu} {\cal F}_3 (\nabla {\cal T})^2
	- \frac{1}{2}{\cal F}_5 G^{\mu\nu} \nabla_\rho {\cal T} \nabla^\rho \Phi
\nn\\
&&	+ \frac{1}{2}{\cal F}_5 \nabla^\mu {\cal T} \nabla^\nu \Phi
	+\frac{1}{2} {\cal F}_5 \nabla^\nu {\cal T} \nabla^\mu \Phi\ ,
\label{einstein}
\ee
where $\nabla^\mu$ is the usual covariant derivative.
Likewise, the dilaton and tachyon equations of motion read, respectively:
\be
0 & = & -2 R f_1 + 8 f_2 (\nabla \Phi)^2
	- 8 f_2' \nabla {\cal T} \cdot \nabla \Phi
	- 8 f_2 \nabla^2 \Phi + (2 f_3 + f_5') (\nabla {\cal T})^2
	+ f_5 \nabla^2 {\cal T} + 4 {\newv}\ ,
\nn\\
&&
\label{dilatoneom}
\\
0 & = & f_1' R + (4 f_2' - 2 f_5)(\nabla \Phi)^2 + f_3' (\nabla {\cal T})^2
	- 4 f_3 \nabla \Phi \cdot \nabla {\cal T} - 2 {\newv}'
	+ f_5 \nabla^2 \Phi + 2 f_3 \nabla^2 {\cal T}\ .
\nn\\
&&
\label{tachyoneom}
\ee

Following the discussion above, we assert that this theory admits 
a solution of the form
\bbb
{\cal T}(X) = \m\sqd \exp{\b X\uu +}\ ,
\qquad
G\ll{\m\n}\us = \eta\ll{\m\n}\ ,
\qquad
\Phi = - q X\uu 0\ .
\eee
First, we note that Eqn.~\rr{teom} (the on-shell condition for the tachyon profile)
sets $\b q = {{2 \sqrt{2}}\over\apr}$.
This demand imposes various conditions on
the functions $f_i({\cal T})$, the simplest of which comes from the lower $(-,-)$ component
of the Einstein equation (\ref{einstein}):
\bbb f\ll 2 = f\ll 1\ .
\een{cons2}
With this condition in place, the $(+,-)$ component 
of Eqn.~\rr{einstein} (minus a multiple of the trace of \rr{einstein}) 
gives
\bbb
f\ll 5 =  4 f'\ll 1\ ,
\een{cons5}
and, assuming this holds, the lower $(+,+)$ component of
Eqn.~\rr{einstein} yields
\bbb
f\ll 3 = - {1\over {\cal T}} f'\ll 1 - f''\ll 1\ .
\een{cons3}
From this we conclude that $f'\ll 1$ must vanish at
${\cal T} = 0$: otherwise $f\ll 3$ would be singular when the tachyon is 
zero.  The rest of the Einstein equation, with both indices 
transverse to the light-cone directions, gives the condition
%
%
\bbb
{\newv}({\cal T}) =  {{D - 26}\over{3\apr}} f\ll 1 + {4\over{\apr}}  {\cal T} f'\ll 1\ .
\een{consv}
We have thus eliminated the four functions $f\ll{2,3,5}$ and ${\newv}({\cal T})$ in
terms of $f\ll 1$.
The remaining two equations, coming from the equations of motion 
for the dilaton and tachyon (in Eqns.~(\ref{dilatoneom}) and (\ref{tachyoneom}) above), 
are satisfied automatically.
We therefore find that, in the presence of a nontrivial tachyon potential
and nonvanishing dilaton, $f_1({\cal T})$ must be nonconstant.  
In other words, the coupling of the 
tachyon to the Einstein term in the effective action must be nontrivial.
It follows that the canonical Einstein frame metric is related to the sigma model
metric via a tachyon-dependent Weyl rescaling.  The Einstein frame metric of the
solution described in Section 3.1 is no longer that of the linear dilaton background,
or that of a homogeneous and isotropic spacetime solution of any kind.
As expected, the backreaction of the tachyon on the background breaks the symmetry 
of the FRW satial slices.

In terms of $f_1$, the final effective action takes the form
\be
S &=& \frac{1}{2\kappa^2}\int d^D x \sqrt{\det G} e^{-2\Phi}\Bigl[
	f_1 R
	+ 4 f_1 (\nabla \Phi)^2
	+ \left( \frac{1}{{\cal T}} f'_1 + f''_1 \right)(\nabla {\cal T})^2
\nn\\
&&
\kern+120pt
	-4 f'_1 \nabla {\cal T} \cdot \nabla \Phi
	-\frac{2}{3\alpha'}(D-26) f_1 - \frac{8}{\alpha'} f'_1 {\cal T}
	\Bigr]\ .
\label{ourfinalaction}
\ee
This form holds for all spacetime dimension $D$.  We can include the critical case 
$D=26$ by considering a tachyon varying in the $X^+$ direction and a dilaton varying in
the $X^-$ direction.   Doing so, we find that the constraints in 
Eqns.~(\ref{cons2},~\ref{cons5},~\ref{cons3},~\ref{consv}) 
hold for the two-derivative action in the critical theory as well.

One might ask whether the equations
of motion generated by our two-derivative
effective action are satisfied by a background in which
a tachyon condenses in a direction {\it other} than a null direction.
This question is addressed in detail in reference \cite{swanson}.

\def\kill{
One exact solution of
critical string theory with non-null tachyon is
the ordinary spacelike Liouville theory with
null linear dilaton: 
\bbb
\Phi &=& - Q (X\uu 0 + X\uu 1)\ ,
\nn\\ \nn \\
{\cal T} &=& \m\sqd \exp{- 2 b X\uu 1}\ , 
\eee
where $Q$ satisfies $Q = b + {1\over{b\apr}}$.
When we repeat the analysis above, we find that the existence
of the spacelike linear dilaton theory implies the relations in 
Eqns.~(\ref{cons2},~\ref{cons5},~\ref{cons3}),
but the form of the potential is modified to
\bbb
{\newv}({\cal T}) = 
	\frac{4}{\alpha'} {\cal T} f'_1 + 2b^2 \left( {\cal T} f'_1 - {\cal T}^2 f''_1 \right)\ .
\een{modifiedconsv}
The spacelike Liouville theory is an exact solution for any $(b,Q)$
satisfying $b + {1\over{b\apr}} = Q$.  
 no functional form
for $f\ll 1({\cal T})$ other than a constant simultaneously 
allows all consistent null {\it and} spacelike Liouville 
theories as exact solutions to the
two-derivative equations of motion.\footnote{The choice 
$f\ll 1({\cal T}) = {\rm const.}$
would set ${\newv}({\cal T}) = 0$ identically.  
A tachyon potential that vanishes identically
makes it hopeless to reproduce tachyon scattering 
amplitudes in the framework of the two-derivative action.} This serves to
remind us that low-energy effective actions are of limited utility
when describing fields with $m\sqd \neq 0$. 
To be sure, the moral is not that the tachyon potential must vanish, 
but that higher derivative terms must be included to obtain a 
consistent effective action.\footnote{Neither potential in 
Eqns.~\rr{consv} or \rr{modifiedconsv}
produces the correct physics in a background with vanishing 
linear dilaton in $D=26$.  In particular, the mass-squared 
would appear as $4/\alpha'$ rather than $-4/\alpha'$.}
}

Working to higher order in $\apr$, there is an enormous freedom to
choose  higher-derivative terms that preserve the null
tachyon background as an exact solution.
The null tachyon solutions satisfy
$(\pp\ll\m \Phi)\cdot (\pp\uu\m {\cal T}) = {2\over{\apr}} {\cal T}$
identically everywhere, so we can add to the action
\bbb
\Delta {\cal L} \equiv
\left( (\pp\ll\m \Phi)\cdot (\pp\uu\m {\cal T}) - {2\over{\apr}} {\cal T}
\right)\sqd 
	\cdot {\cal F}(G\ll{\m\n}, \Phi,{\cal T})\ , 
\eee
where ${\cal F}$ is an arbitrary functional of the spacetime
fields and their derivatives.   (The modified action will
still generate equations of motion that are
satisfied by ${\cal T} = \m\sqd \exp{\b X\uu +},~ \Phi = - q
X\uu 0,~ G\ll{\m\n} = \eta\ll{\m\n}$.) 
It would be interesting to see whether higher-derivative actions can be
found that admit {\it all} known exact solutions to the
equations of motion.

\section{Discussion and conclusions}
We have shown that timelike linear dilaton theories provide a simple setting for the study of
time-dependent backgrounds in string theory.  Quintessent theories
arise naturally from string theory in supercritical dimensions $(D > D_{\rm crit})$.
As one might expect, the equation of state $w$ of 
these cosmologies depends on the number
of spacetime dimensions $D$. 
The tree level potential of supercritical string
theory gives rise to
the threshold value $w_{\rm crit} = -(D-3)/(D-1)$
between accelerating and non-accelerating
cosmologies.  For
$w = w\ll{\rm crit}$,
the universe is globally conformally equivalent to Minkowski space.
We have also described exact solutions,
which behave as a bubble of nothing
in these backgrounds.  We have used the existence
of this class of solutions to constrain the effective action of string theory.

\heading{Relation to recent work}

Recently, a paper \cite{takayanagi} appeared that 
examines a bubble solution related to the one presented in this paper.  The authors
consider a solution describing tachyon condensation 
along a spacelike direction (in critical spacetime dimension $D=D_{\rm crit}$)
with a lightlike linear dilaton:
\bbb
\Phi =  -Q (X\uu 0 + X\uu 1)\ .
\eee
The tachyon profile is
\bbb
{\cal T} = \m \sqd \exp{ - 2 b X\uu 1}\ ,
\een{tachpert}
with $b + {1\over \alpha' b} = Q$.  They boost the solution
so that the dilaton and tachyon become
\bbb
\Phi &=& -{Q\over \g} (\tilde{X}\uu 0 + \tilde{X}\uu 1)\ ,
\nn\\ 
&&
\nn\\
{\cal T} &=& \m\sqd \exp{b\, \g\, (\tilde{X}\uu 0 - \tilde{X}\uu 1)
- {b\over{\g}}\, (\tilde{X}\uu 0 + \tilde{X}\uu 1) }\ .
\eee
The authors then take a formal limit $\g\to \infty$ and
describe the resulting theory as a simple example of
a spacetime boundary in a sigma model,
dropping the
$\g\uu{-1}$ term in
the exponent of the Liouville potential,
and setting $\Phi$ to a constant.
The resulting worldsheet potential is identical
to our own, but only formally:  the theory of \cite{takayanagi}
differs from ours at the level of observable quantities, such as operator
dimensions and OPEs.

The apparent simplicity of the limit $\g\to \infty$ is deceptive.
The most evident difficulty
is that the tachyon coupling to the string worldsheet
(see Eqn.~\rr{tachpert}) is not marginal unless one retains the
$\g\uu{-1}$ term in the Liouville exponent.  Indeed,
the anomalous dimension of the exponential is always
equal to $- \apr b\sqd$.  In particular, it is of
order $\g\uu 0$.  This anomalous dimension
is an invariant, which cannot be boosted away. 

Because the tachyon gradient appears
formally null, 
a literal application of the approach of
\cite{takayanagi}
would treat the anomalous dimension as vanishing,
and the same would hold for the dilaton contribution
to the dimension
of the exponential (which is also of order $\g\uu 0$).
For a consistent treatment of the Lagrangian in
\cite{takayanagi}, one needs to retain the $\g\uu{-1}$
terms in the dilaton and in the Liouville exponent.
Loop corrections to the worldsheet theory are
always of order unity, though some amplitudes in 
the theory may still be solvable by virtue of the
special properties of Liouville theories.

By contrast, the tachyon exponent in our model
is null from the beginning, and its
anomalous dimension truly vanishes.  The worldsheet
theory is solvable not because of any
special properties of Liouville theory, but because 
all loop corrections to semiclassical
amplitudes vanish.

Dropping the $\g\uu{-1}$ terms for
large boost also amounts to setting
the terms $\pp\ll\m\pp\uu \m  {\cal T}$, $(\pp\ll\m {\cal T})\sqd$
and $(\pp\ll\m \Phi)(\pp\uu\m {\cal T})$ 
to zero in the equations of motion for the 
spacetime fields.  To analyze the constraints on the
effective action consistently, one must restore them.
Doing so, one recovers the constraints 
(\ref{cons2},~\ref{cons5},~\ref{cons3}).
Additionally, if one makes an ansatz that the tachyon dependence of the Einstein term
is trivial, as does \cite{takayanagi}, one finds that the 
tachyon potential must vanish identically for all values of ${\cal T}$.
Allowing the Einstein term to depend nontrivially on the
tachyon, one can accommodate a nontrivial tachyon potential,
for a particular form of $f_1(\ct)$.  This result is reported in
\cite{swanson}.  In particular, ${\newv}({\cal T})$ need {\it not} 
vanish as ${\cal T}\to\infty$.

\heading{Outlook}

Theories describing null
tachyon condensation may profitably be explored
and generalized further.  In particular, one can perturb the worldsheet theory of
the supercritical string with an operator of the form $\exp{\b X\uu +} {\cal O}$,
where ${\cal O}$ is a relevant operator of weight $h$, and $\b q = (1 - h) 2\sqrt{2} / \apr$.
The lightlike Liouville dressing of the operator still behaves classically,
so we expect that such a theory should be just as solvable as the nonconformal 2D 
theory defined by perturbing with the undressed operator ${\cal O}$.  In some
cases ${\cal O}$ can induce a renormalization group flow to a nontrivial
CFT with vanishing vacuum energy, either as a result of worldsheet supersymmetry or by fine-tuning
the vacuum energy by hand.  When ${\cal O}$ has this property, the 
corresponding string theory can be interpreted as a dynamical transition between
two vacua with different total numbers of spacetime dimensions.
We plan to address these and related issues in future studies.


\section*{Acknowledgments}
We thank Nemanja Kaloper, David Kutasov, Joseph Polchinski, Savdeep Sethi, Leonard
Susskind and Mark Trodden for valuable conversations.
S.H.~is the D.~E.~Shaw \& Co.,~L.~P.~Member
at the Institute for Advanced Study.
S.H.~is also supported by U.S.~Department of Energy grant DE-FG02-90ER40542. 
I.S.~is the Marvin L.~Goldberger Member
at the Institute for Advanced Study, and is supported additionally
by U.S.~National Science Foundation grant PHY-0503584. 

\bibliographystyle{utcaps}
\bibliography{dimchange}

\end{document}